# Combined parameterization of material distribution and surface mesh for stiffener layout optimization of complex surfaces

Weihong Zhang* and Shengqi Feng

*State IJR Center of Aerospace Design and Additive Manufacturing,*
*Northwestern Polytechnical University, 710072 Xi'an, Shaanxi, China*
*Corresponding author zhangwh@nwpu.edu.cn

**Abstract**

Stiffener layout optimization of complex surfaces is fulfilled within the framework of topology optimization. A combined parameterization method is developed in two aspects. One is to parameterize the material distribution of the stiffener layout by means of B-spline. The other is to build the mapping relationship from the known 3D surface mesh of the thin-walled structure to its parametric domain by means of mesh parameterization. The influence of mesh parameterization upon the stiffener layout is discussed to reveal the matching issue of the combined parameterization. 3D complex surfaces represented by the triangular mesh can be dealt with even though analytical parametric equations are not available. Some numerical examples are solved to demonstrate the direct advantage and effectiveness of the proposed method.

## 1 Introduction

Stiffener layout optimization of complex surfaces is very popular in aeronautical and aerospace industries. For this kind of problem, topology optimization method is recognized as an efficient tool and being extended to achieve high structural performance and lightweight designs. Earlier important works can be outlined after a comprehensive literature review. Typically, Cheng and Olhoff (1981, 1982) studied stiffener layout to optimize the thickness distribution onto the rectangular and axisymmetric plates. Afonso et al. (2005) presented an integration procedure of sizing, shape and topology optimization for stiffened thin-walled structures by means of homogenization method. Tian et al. (2020) developed a data-driven modeling and stiffener optimization method for undevelopable curved surfaces. Ding and Yamazaki (2004), Ji et al. (2014) developed the so-called bionic growth method to optimize the stiffener layout with the inspiration from branching tree systems in nature. Liang et al. (2020), Hao et al. (2016), Wang et al. (2017, 2019, 2020) focused on stiffener layout optimization for buckling problems. Besides, much effort was made on the cutout design on the thin-walled structures. Wang and Zhang (2012) used parametric mapping to realize hole shape optimization on the surface. Kiendl et al. (2014), Kang et al. (2016), Seo et al. (2010) developed isogeometric methods for shape and topology optimization of thin-walled structures.

The extension of B-spline parameterization method to topology optimization was initially developed by Qian (2013) and can be considered as a parameterization of the existing density method with element density variables replaced with control points. In our previous work (Feng et al. 2021), this method was further extended for stiffener layout optimization of complex surfaces, but the analytical parametric equation should be absolutely available in advance for the complex surface. The reason is that the determination of the parametric space required for the B-spline parameterization of the stiffener layout depends upon the parametric equation of the complex surface. This requirement is, however, a great challenge that strongly limits the application of our previous method (Feng et al. 2021). For example, when the complex surface of a thin-walled structure is made of joint surface patches or represented approximately in form of 3D meshed surface, it is difficult and even impossible to obtain the parametric equation of the complex surface.

In view of the above considerations, the combined parameterization of material distribution and surface mesh is proposed. In our case, as the 3D surface is numerically represented in form of the triangular mesh due to its geometric complexity, the mesh parameterization aims at establishing a one-to-one mapping from the 3D mesh surface to the 2D parametric domain. Historically, mesh parameterization was firstly developed in the community of computer graphics to map textures onto surfaces (Bennis et al. 1991; Maillot et al. 1993). Later, the method was improved by Floater (1997) to represent each interior vertex of the mesh as a convex combination of its neighboring vertices based on the barycentric mapping theorem of Tutte (1960). Lévy et al. (2002) further developed the least-square approximation to achieve a continuous conformal mapping. A local/global approach was proposed by Liu et al. (2008) to parameterize the mesh in a shape-preserving manner and has gradually become a ubiquitous tool for many mesh processing applications (Sheffer et al. 2006; Hormann et al. 2007).

The outline of the paper is as follows: B-spline parameterization method and mesh parameterization are presented in Section 2 and Section 3, respectively. The matched combination of both parameterizations is discussed in Section 4. Stiffener layout optimization and sensitivity analysis are presented in Section 5. Several numerical examples are solved in Section 6. Conclusions are finally drawn out in Section 7.

## 2 B-spline parameterization for stiffener layout modeling

At the early stage, B-spline parameterization was a commonly used function for data fitting and computer-aided modeling. Recently, it is combined with the density method in topology optimization where design variables are directly associated with control points instead of element pseudo-densities (Qian 2013). Here, B-spline parameterization is further employed to model the material distribution as a continuous B-spline field of stiffener layout.

Mathematically, a univariant B-spline basis function $N_{i,p}(\xi)$ is based on a knot vector $\boldsymbol{I} = \{\xi_1, \xi_2, \ldots, \xi_{n+p+1}\}$ consisting of a non-decreasing sequence of knots. It is defined recursively as

$$\begin{cases} N_{i,0}(\xi) = \begin{cases} 1, & \xi_i \leq \xi < \xi_{i+1} \\ 0, & \text{otherwise} \end{cases}, \\ N_{i,p}(\xi) = \dfrac{\xi - \xi_i}{\xi_{i+p} - \xi_i} N_{i,p-1}(\xi) + \dfrac{\xi_{i+p+1} - \xi}{\xi_{i+p+1} - \xi_{i+1}} N_{i+1,p-1}(\xi) \ i = 0,1,\ldots,n-1 \\ \text{assuming } \dfrac{0}{0} = 0 \end{cases} \quad (1)$$

where $p$ represents the degree of basis function, $n$ represents the number of basis functions and control points.

Based on the above definition, the stiffener layout can be parameterized by a bivariant B-spline height field in the following form

$$h = \sum_{i=0}^{n-1} \sum_{j=0}^{m-1} N_{i,p}(\xi) N_{j,q}(\eta) \bar{h}_{ij}, \quad 0 \leq \xi, \eta \leq 1 \quad (2)$$

where $\xi$ and $\eta$ denote intrinsic coordinates of the parametric domain; $N_{i,p}$ and $N_{j,q}$ are the basis functions of degree $p$ in $\xi$-direction and degree $q$ in $\eta$-direction, respectively. $\bar{h}_{ij} \in [0, H_{max}]$ are height design variables associated with B-spline control points with their numbers $n$ and $m$ in $\xi$- and $\eta$-directions, respectively.

Based on the B-spline height field $h$, the pseudo-density of each finite element is easily evaluated according to its position (Liu et al. 2015).

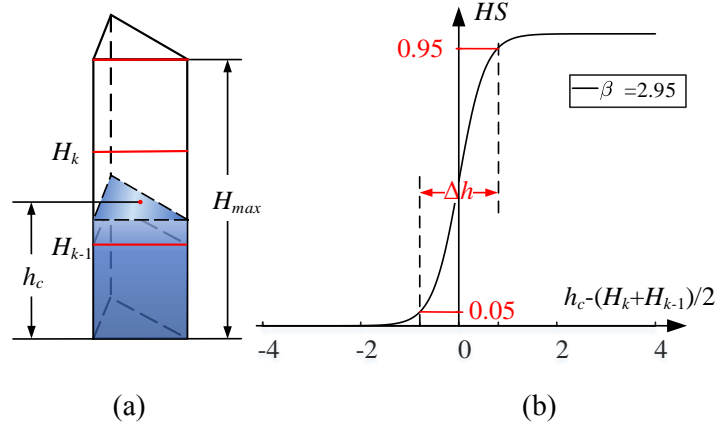

(a) (b)

Fig. 1 Calculation of pseudo-density for boundary elements. (a) Calculation of pseudo-density with height ratio. (b) Heaviside function and its transition interval $\Delta h$.

Fig. 1(a) shows how the pseudo-density is calculated for boundary element $k$. A Heaviside function $HS$ shown in Fig. 1(b) is introduced to project the element pseudo-density as

$$\rho_k = HS(h_c - (H_k + H_{k-1})/2) = \frac{e^{\beta(h_c - (H_k + H_{k-1})/2)}}{1 + e^{\beta(h_c - (H_k + H_{k-1})/2)}} \quad (3)$$

Here, $h_c$ is the height value at the centroid of element $k$. $\beta > 0$ is the internal parameter to control the steepness of the projection.

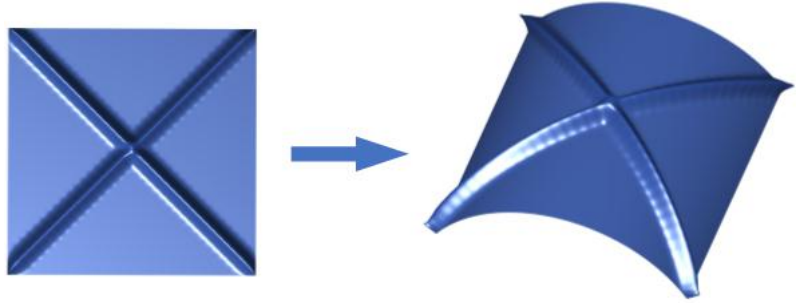

Fig. 2 Illustration of parametric mapping.

For a thin-walled structure with curved surface, Eq. (2) corresponds to the map of stiffener height field from the parametric domain to the physical space, as shown in Fig. 2. This map can easily be obtained whenever the following mapping relation exists for the concerned surface.

$$f: \quad (\xi, \eta) \to (x, y, z) \tag{4}$$

In fact, the one-to-one mapping between the parametric domain and the physical domain is the key to the application of the B-spline parameterization method. For 3D simple and regular surfaces, the above mapping can analytically be obtained because their parametrical equations of closed-form are available. However, for complex surfaces studied in this work, they are often approximated in the discretized form of triangular meshes. In this case, the mapping relationship should be constructed numerically by means of mesh parameterization presented below.

## 3  Mesh parameterization for 3D meshed surface

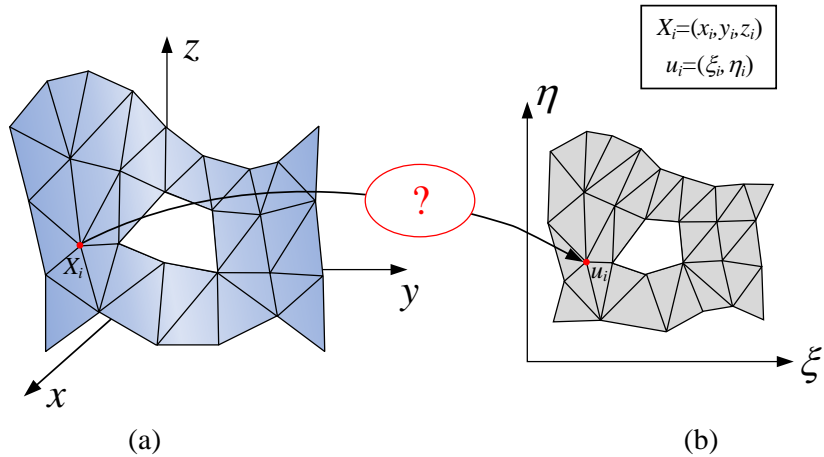

(a)                                (b)

Fig. 3 Schematic diagram of mesh parameterization. (a) 3D meshed surface. (b) Mesh parameterization.

In engineering practice, 3D triangular meshing is very popular to model complex surfaces. Mesh parameterization is to compute the one-to-one mapping from a 3D meshed surface in the Cartesian coordinate system to the parametric domain. According to the feature of parametric domain, mesh parameterization methods can be classified into planar parameterization, spherical parameterization, base mesh parameterization, etc. Fig. 3 illustrates the planar parameterization method of a 3D meshed surface.

Therefore, the mapping represents the relationship between 3D coordinates $(x, y, z)$ and the parametric coordinates $(\xi, \eta)$ of the mesh. For node $i$, the mapping corresponds to

$$f: (x_i, y_i, z_i) \to (\xi_i, \eta_i) \tag{5}$$

However, whenever a triangular meshed surface is given, a variety of parametric planar meshes can be obtained. The ideal solution is to minimize the distortion of the parametric planar mesh without overlapping. To this end, measures should be developed for the quality control of mesh parameterization.

To make ease the discussion, the 3D meshed surface can be classified according to its topological information. A multi-boundary surface means that holes exist inside the surface. A single-boundary surface means that no inner holes exist, as seen in Fig. 4(a). A closed surface such as spherical surface means that no boundary exists. Below are presented typical mesh parameterization methods.

### 3.1 Planar parameterization method

- **Convex combination approach**

Planar parameterization is the most studied and the simplest method that flattens a spatial triangle mesh into a planar one as uniformly as possible. Suppose $X_1, ..., X_g$ denote interior nodes of the triangle mesh shown in Fig. 4. $X_{g+1}, ..., X_G$ denote the boundary points. The basic idea of the convex combination is to fix the boundary points a priori in the parametric domain and then determine each interior point through the weighted convex combination of its neighboring points.

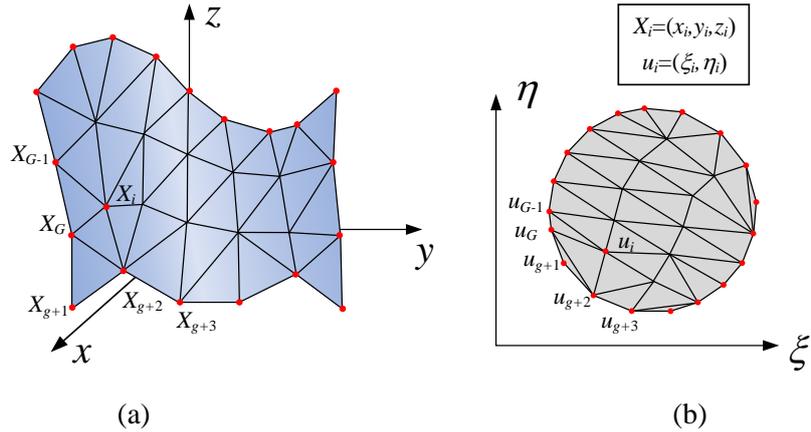

(a)           (b)

Fig. 4 Convex combination approach with circular boundary in the parametric domain.

To do this, boundary points are first mapped counterclockwise onto a circle (or square) according to the length proportions. The parametric coordinates of boundary point $u_i$ are then calculated as

$$\begin{cases} \xi_i = (\sin(\dfrac{\sum\limits_{j=g+1}^{i}\|X_{j+1}-X_j\|}{\sum\limits_{j=g+1}^{G}\|X_{j+1}-X_j\|}\cdot 2\pi)+1)/2 \\ \\ \eta_i = (\cos(\dfrac{\sum\limits_{j=g+1}^{i}\|X_{j+1}-X_j\|}{\sum\limits_{j=g+1}^{G}\|X_{j+1}-X_j\|}\cdot 2\pi)+1)/2 \quad i=g+1,...,G \\ \\ \text{assuming} \quad X_{G+1}=X_{g+1} \end{cases} \quad (6)$$

Second, the parametric coordinate of each interior point $i$ is calculated by means of convex combination

$$u_i = \sum_{j \in G(i)} \lambda_{ij} u_j \quad i=1,\ldots,g \tag{7}$$

Here, weight $\lambda_{ij}$ depends upon neighboring points of point $i$ and satisfies the following condition.

$$\begin{cases} 0 < \lambda_{ij} < 1, \sum_{j=1}^{G(i)} \lambda_{ij} = 1 & i=1,\ldots,g \\ 0 & i=g+1,\ldots,G \end{cases} \tag{8}$$

where $j \in G(i)$ is the number of neighboring points connected to point $i$. For example, $\lambda_{ij}$ can be assigned to locate the internal point as the center of gravity.

$$\lambda_{ij} = \begin{cases} \dfrac{1}{G(i)} & i=1,\ldots,g \\ 0 & i=g+1,\ldots,G \end{cases} \tag{9}$$

All internal points can thus be determined in terms of known boundary points $u_{g+1}, \ldots, u_G$ by solving the linear system of Eq. (7). Clearly, this approach is only applicable to single-boundary surfaces because meshes are overlapped for multi-boundary problems. Moreover, boundary points mapped onto a circle might produce a significant mesh distortion in the parametric domain.

- **Energy minimization approach**

This kind of approach defines such an energy-based objective function whose extrema are sought to determine the parametric coordinates. The harmonic parameterization approach (Eck et al. 1995) is a typical one that treats each mesh edge as an elastic spring so that the total energy of the surface mesh reads

$$\Phi(u) = \frac{1}{2} \sum_{(i,j) \in S} k_{ij} \|u_i - u_j\|^2 \tag{10}$$

where $S$ denotes the set of all edges; $k_{ij}$ represents the elasticity coefficient of the triangle edge between point $i$ and point $j$. When the total energy is minimal, the extreme value corresponds to

$$\frac{\partial \Phi(u)}{\partial u} = \sum_{(i,j) \in S} k_{ij}(u_i - u_j) = 0 \tag{11}$$

Similarly, this equation is solved by substituting the parametric coordinates of the boundary points obtained from Eq. (6).

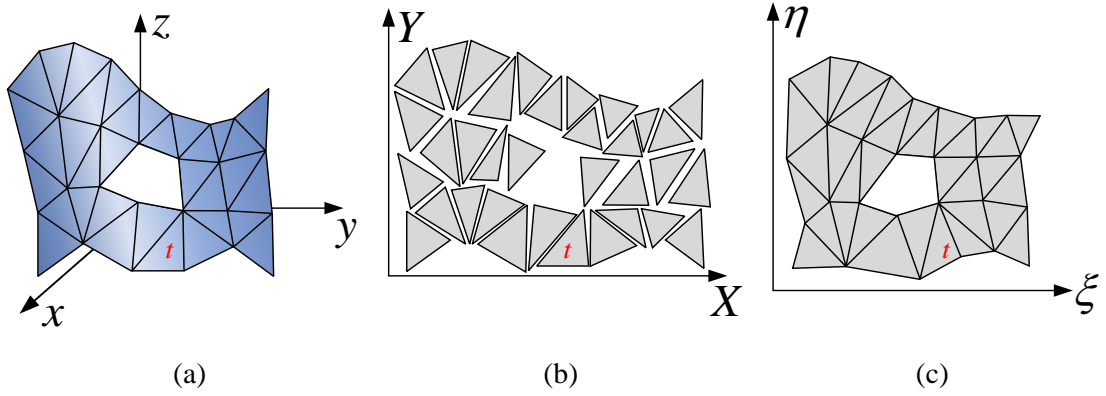

(a)              (b)              (c)

Fig. 5 ARAP approach. (a) Original 3D mesh. (b) Flattening each triangular mesh independently onto the plane. (c) Mesh parameterization with local alignment of flattened triangles.

More importantly, the so-called ARAP (As-Rigid-As-Possible) is the mainstream energy minimization approach in commercial software and aims at producing the least distortion (Liu et al. 2008). This approach first transforms each original triangular mesh to a planar one in a shape-preserving manner through coordinate transformation. Fig. 5 illustrates planar triangles. All triangular meshes are first transformed to the $XY$ plane by mutually independent congruent transformations and suppose triangle $t$ has its nodal coordinates $X_t = \{X_{t1}, X_{t2}, X_{t3}\}$ in the $XY$ plane. The parametric coordinates $u_t^0 = \{u_{t1}^0, u_{t2}^0, u_{t3}^0\}$ can correspondingly be calculated based on the convex combination approach mentioned above and adopted as initial values. For each triangle after coordinate transformation, the mapping from 2D coordinates $X_t$ to parametric coordinates $u_t$ can be characterized by the Jacobian matrix $J_t$.

$$J_t(u) = \frac{\partial(\xi, \eta)}{\partial(X, Y)} = \begin{bmatrix} \dfrac{\partial \xi}{\partial X} & \dfrac{\partial \xi}{\partial Y} \\ \dfrac{\partial \eta}{\partial X} & \dfrac{\partial \eta}{\partial Y} \end{bmatrix} \quad (12)$$

An auxiliary linear transformation matrix $L_t$ of dimension $2 \times 2$ is then assigned to triangle $t$. This matrix holds the following form to allow the rotation and similarity transformations.

$$L_t = \left\{ \begin{pmatrix} a_t & b_t \\ -b_t & a_t \end{pmatrix} : a_t, b_t \in \mathbf{R} \right\} \quad (13)$$

An energy measure involving $u$ and $L_t$ is thus defined as

$$\Phi(u, L_t) = \sum_{t=1}^{T} A_t \|J_t - L_t\|_F^2 \quad (14)$$

where $\|\cdot\|_F$ is the Frobenius norm; $A_t$ is the area of triangle $t$; $T$ is the total number of triangles. Following Pinkall and Polthier (1993), Eq. (14) can be rewritten in terms of the mesh half-edges as

$$\Phi(u, L_t) = \frac{1}{2} \sum_{(i,j) \in he} \cot(\theta_{ij}) \|(u_i - u_j) - L_{t(i,j)}(X_i - X_j)\|^2 \quad (15)$$

where $he$ is the set of mesh half-edges. $t(i, j)$ is the triangle containing the half-edge $(i, j)$, and $\theta_{ij}$ is the angle opposite $(i, j)$ in $t(i, j)$. Hence, the parametric mesh is obtained by solving the energy minimization problem

Find: $u, a_t, b_t$ $(t = 0, 1, \cdots, T)$

$$\min : \Phi(u, a_t, b_t) = \frac{1}{2} \sum_{(i,j) \in he} \cot(\theta_{ij}) \left\| (u_i - u_j) - \begin{pmatrix} a_{t(i,j)} & b_{t(i,j)} \\ -b_{t(i,j)} & a_{t(i,j)} \end{pmatrix} (X_i - X_j) \right\|^2 \quad (16)$$

s.t. $a_t, b_t \in \mathbb{R}$

This implies that the parametric mesh should be as close as possible to the original mesh through rotation and similarity transformation. The numerical solution can be efficiently carried out. First, use the initial Jacobian matrix $J_t^0$ obtained in terms of $X_t$ and $u_t^0$ to solve $L_t^0$ via the singular value decomposition (SVD). For fixed $L_t^0$, the minimum $u^1$ can be found by setting $\frac{\partial \Phi(u, L_t)}{\partial u} = 0$ so that

$$\sum_{j \in G(i)} (\cot(\theta_{ij}) - \cot(\theta_{ij}))(u_i - u_j) = \sum_{j \in G(i)} (\cot(\theta_{ij}) L_{t(i,j)} + \cot(\theta_{ji}) L_{t(j,i)})(X_i - X_j) \quad i=1,\ldots, G \quad (17)$$

The new Jacobian matrix $J_t^1$ and the corresponding $L_t^1$ can be iterated by updating $u_t^1$. Here, the superscript represents the iteration number. $u_t^0$ represents the initial value of the iteration. $u_i$ and $u_j$ represent the parameter coordinates of the $i$-th and $j$-th vertices of the triangular mesh. The optimal $u$ and $L_t$ are obtained by repeating the above iterative steps until the energy change between two successive steps is less than the prescribed tolerance. Note that this approach can be applied to multi-boundary surfaces.

Clearly, both the convex combination and the energy minimization approaches are dedicated to the mesh parameterization of bounded surfaces. Table 1 recapitulates both approaches for single and multi-boundary surfaces.

| 3D meshed surface | Convex combination approach | Energy minimization approach (ARAP) |
|---|---|---|
| 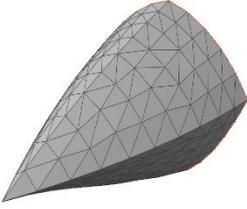 Single-boundary surface | 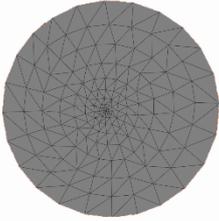 | 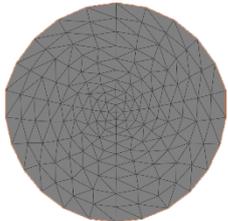 |
| 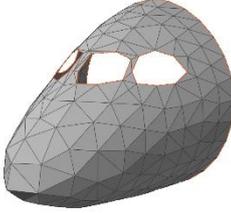 Multi-boundary surface | 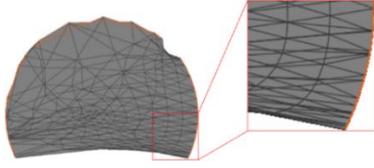 | 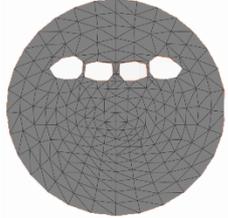 |
| Features | Large mesh distortion and overlapping for multi-boundary surfaces | Small mesh distortion for both single and multi-boundary surfaces |

Table 1 Comparison of convex combination approach and energy minimization approach.

- **Cutting approach**

The cutting approach is devised for closed surfaces that are not suitable for the direct application of planar parameterization. Fig. 6 depicts a closed surface cut into a single boundary surface with introduction of additional 3D points along the cutting. In this manner, the energy minimization approach described above can be extended directly.

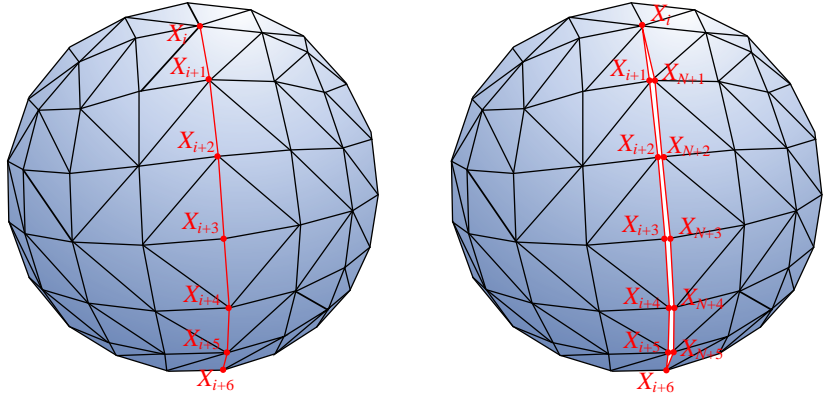

Fig. 6 Seam cutting approach.

However, the cutting will inevitably introduce surface discontinuities into the parameterization, a delicate balance must be made between the conflicting goals of small distortion and short cutting. For this reason, the multi-chart segmentations approach (Sander et al. 2003) and the seam cutting approach (Sorkine et al. 2002) were proposed. The former is to partition complex surfaces into several charts and flatten them onto the parametric plane. It is hoped that the distortion of each parameterized chart is as small as possible while keeping the cutting boundaries short enough and the number of charts small enough. The latter is used for closed surfaces or surfaces with large distortion after mesh parameterization. Multiple cuttings, i.e., seams are introduced into the surface but not cut into separate charts. Comparatively, the seam cutting approach produces a shorter cutting boundary than the multi-chart segmentations approach.

Fig. 7 depicts a spherical model preprocessed by two cutting approaches. Fig. 7(a) and (b) show the sphere surface cut into two and four charts and flattened onto the plane, respectively. Fig. 7(c) plots the parameterized mesh by the seam cutting approach. As is seen, the multi-chart segmentations approach produces the unfolding result with a longer cutting boundary than the seam cutting approach, but produces a smaller mesh distortion.

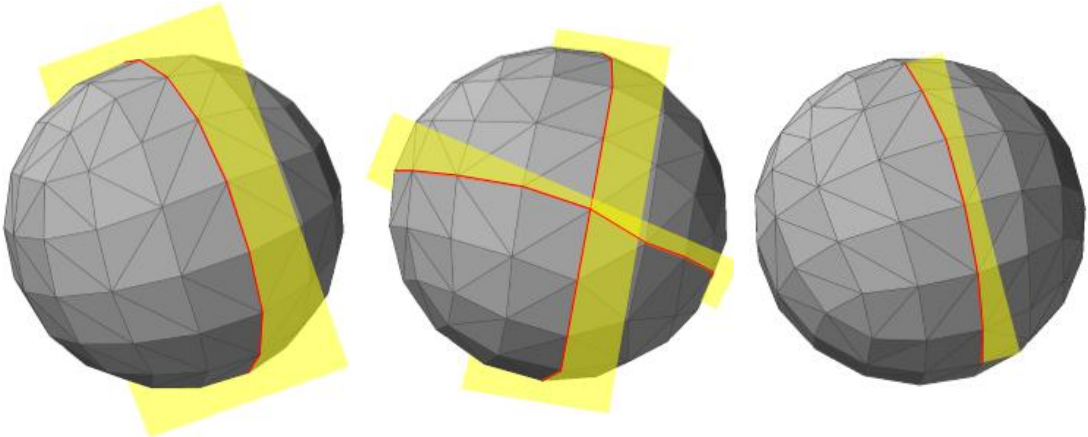

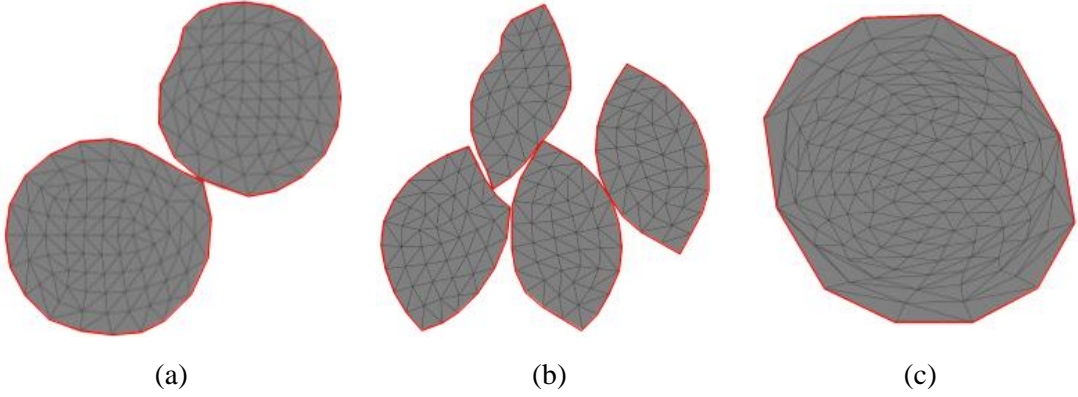

      (a)                    (b)                    (c)

Fig. 7 Examples of two cutting approaches. (a) and (b) Multi-chart segmentations approach with different numbers of charts. (c) The seam cutting approach.

### 3.2 Spherical parameterization method

This method concerns the mesh parameterization for closed surfaces in order to avoid cutting discontinuities caused by the cutting approach. The closed surface is mapped to a closed spherical parametric domain. Fig. 8 shows the spherical parameterization that seamlessly unfolds the closed surface of a submarine model with small distortion. In addition, other parameterization methods such as column parameterization and base mesh parameterization also exist (Sheffer et al. 2006; Hormann et al. 2007). In this paper, the ARAP energy minimization approach and the cutting approaches are applied to realize the one-to-one mapping.

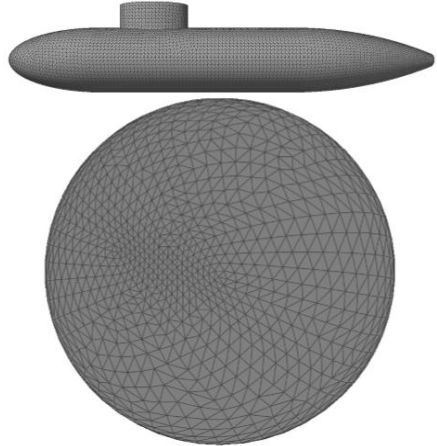

Fig. 8 Illustration of spherical parameterization method.

## 4 Matching issue between mesh parameterization and B-spline parameterization

Mesh parameterization makes it possible to represent the 3D meshed surface onto a parametric domain. It is a vital step to bridge the meshed surface model, structure performance analysis and stiffener layout optimization. In nature, this is a kind of numerical parameterization because the triangular mesh is used as a discrete approximation to the analytical surface.

In this sense, any modification and design of a geometric feature in the parametric domain should be well preserved on the 3D meshed surface after mapping. The B-spline parameterization should produce a uniform distribution of control points dominating the stiffener height field in accordance with the mesh parameterization to reduce the mesh distortion as small as possible. In other words, although design variables are related to B-spline control points independently of mesh parameterization, the mesh parameterization should match B-spline parameterization for stiffener layout optimization. Here, the influence of mesh parameterization on B-spline parameterization is firstly studied. The quality control of mesh parameterization is then presented.

**4.1 The influence of mesh parameterization on the B-spline parameterization**

Fig. 9 shows an example of the combined parameterization results for discussions. Suppose the parameterized mesh has initially a non-uniform distribution, while the B-spline control points have a uniform distribution, as shown in Fig. 9(a). These control points dominate the physical field represented by colors, which can be considered here as a temperature field. This mismatch results in an overwhelmingly yellow color over the 3D meshed surface, while the blue color represented by nearly half of the total control points in the parametric domain occupies only one mesh in the physical domain. In contrast, Fig. 9(b) shows the uniform mesh parameterized by the convex combination approach in accordance with the uniform distribution of control points by the B-spline parameterization. In this case, the color distribution in both the physical and parametric domain exhibits a good agreement. In Fig. 9(c), control points are locally intensified in number. They accord with the non-uniformity of the parameterized mesh to ensure a matched color distribution in both domains.

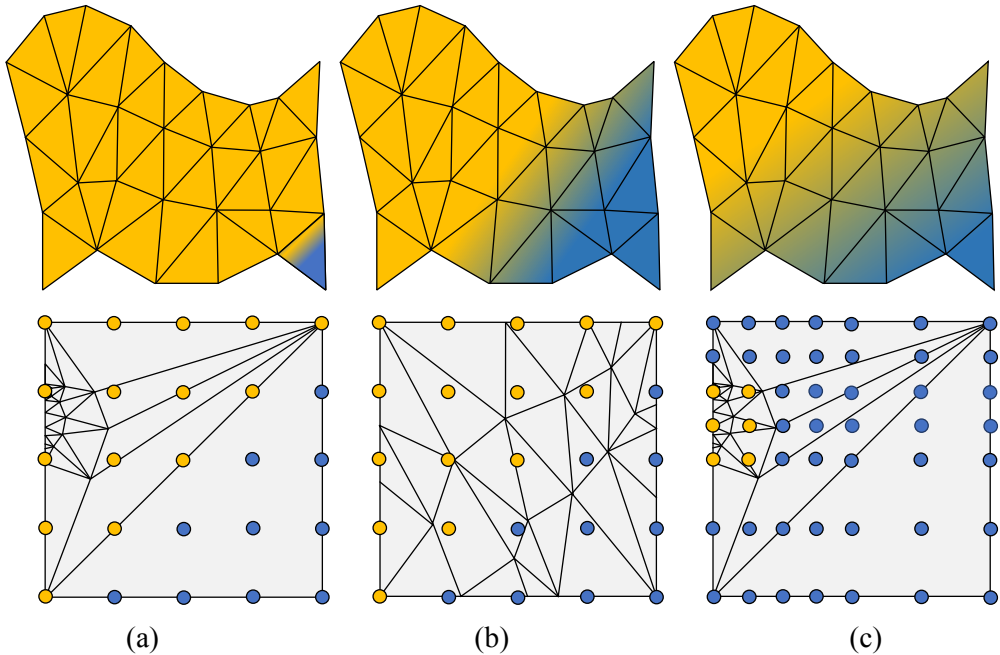

(a)　　　　　　　　　(b)　　　　　　　　　(c)

Fig. 9 Effects of mesh parameterization and B-spline parameterization. (a) Uniform control points but non-uniform mesh parameterization. (b) Uniform control points and mesh parameterization. (c) Locally intensified control points and non-uniform mesh parameterization.

**4.2 Evaluation of mesh parameterization quality**

As is seen, mesh parameterization results might be different even for the same meshed surface. Therefore, the quality of the mesh parameterization should be properly controlled. Sheffer and Sturler (2000) proposed to use the area or angle distortions of triangles as two metrics.

$$D_{area} = \sum_{t=1}^{T} \left( \frac{A_t}{\sum_{i=1}^{T} A_i} - \frac{A_t^*}{\sum_{i=1}^{T} A_i} \right)^2 \quad (18)$$

$$D_{angle} = \sum_{t=1}^{T} \left( \sum_{i=1}^{3} \left( \frac{S_{ti}}{2\pi} - \frac{S_{ti}^*}{2\pi} \right)^2 \right)$$

where $T$ denotes the total number of triangles; $A_t$ and $A_t^*$ represent the area of the $t$-th triangle on the 3D surface and the parametric domain, respectively; $S_{ti}$ and $S_{ti}^*$ denote the degree of the $i$-th angle of the $t$-th triangular elements on the 3D surface and parametric domain, respectively. Both metrics represent an overall measure of distortion and do not reflect the local mesh distortion.

Sander et al. (2001) proposed the geometric-stretch-based metric. The Jacobian matrix $J_t$ is first solved. Thereafter, the maximum eigenvalue $\Gamma_t$ and minimum eigenvalue $\gamma_t$ of $J_t$ are calculated. For the $t$-th triangle, two stretching measures are defined as

$$D_t^2 = \sqrt{(\Gamma_t^2 + \gamma_t^2)/2}$$
$$D_t^\infty = \Gamma_t \quad (19)$$

where $D_t^2$ denotes the root mean square of the stretch in each direction of the $t$-th triangular mesh, and $D_t^\infty$ is the maximum stretch of the $t$-th triangular mesh. For the entire mesh, both stretching measures can be expressed as

$$D_{global}^2 = \sqrt{\frac{\sum_{t=1}^{T}(D_t^2)^2 A_t}{\sum_{i=1}^{T} A_i}} \quad (20)$$

$$D_{global}^\infty = \max_{t \in T} D_t^\infty$$

This method can reflect both the local and global mesh distortions and can be regarded as commonly used metrics.

In this work, $D_t^2$ is used as a metric parameter to measure the mesh parameterization. It is normalized in accordance with the scaling of the parametric domain.

$$D_t = D_t^2 \sqrt{\frac{\sum_{i=1}^{T} A_i}{\sum_{i=1}^{T} A_i^*}} \quad (21)$$

When parameter $D_t = 1$, the triangle on the 3D meshed surface is neither stretched nor compressed after mesh parameterization. If parameter $D_t > 1$, the triangle on the 3D meshed surface is stretched after mesh parameterization. Practically, an allowable range [0.5, 2] is defined. This range can be interpreted as the parameterized triangle area with 1/2 times smaller or 2 times larger than the original triangle area. If the value is beyond the range, the mesh parameterization results are not considered to match the B-spline parameterization for stiffener layout optimization. The cutting approach should be applied for the surface partition. Furthermore, if we are concerned with a closed surface, the cutting approach is mandatory for preprocessing. The seam cutting approach is considered as the preferred option due to the relatively short seam introduced. Since the closed surface is still complex after the seam cutting, the mesh distortion is inevitably large after the planar parameterization. Therefore, the allowable range for the seam cutting approach is relaxed to [0.25, 4]. If the distortion still exceeds the range after using the seam cutting approach, further cutting with the multi-chart segmentations approach is needed.

## 5 Stiffener layout optimization over 3D meshed surface

### 5.1 Problem formulation

Based on the concept presented in Section 2, the stiffener layout is modeled as a B-spline field whose control points act as design variables to define the stiffener height parameters $h$. Here, the minimization of structural compliance corresponds to

$$\text{Find}: \bar{h}_{ij} \ (i=0,1,\cdots,n\text{-}1;\ j=0,1,\cdots,m\text{-}1)$$

$$\min: C(\bar{h}) = \frac{1}{2}\mathbf{U}^{\mathrm{T}}\mathbf{K}\mathbf{U} \tag{22}$$

$$\text{s.t.} \begin{cases} \mathbf{K}(\bar{h})\mathbf{U} = \mathbf{F} \\ g(\bar{h}) \leq 0 \\ 0 \leq \bar{h}_{ij} \leq H_{max} \end{cases}$$

where $C$ is the structural compliance. $\mathbf{K}$ and $\mathbf{U}$ are the global stiffness matrix and displacement field for the stiffened thin-walled structure, respectively; $\mathbf{F}$ is the force vector which is assumed to be irrelevant to design variables; $g(\bar{h})$ denotes the volume constraint to solid materials.

### 5.2 Sensitivity analysis

The sensitivity of the objective function $C$ with respect to $\bar{h}_{ij}$ can be derived as

$$\frac{\partial C}{\partial \bar{h}_{ij}} = \frac{\partial \mathbf{F}}{\partial \bar{h}_{ij}} - \frac{1}{2}\mathbf{U}^{\mathrm{T}}\frac{\partial \mathbf{K}}{\partial \bar{h}_{ij}}\mathbf{U} = -\frac{1}{2}\mathbf{U}^{\mathrm{T}}\frac{\partial \mathbf{K}}{\partial \bar{h}_{ij}}\mathbf{U} \tag{23}$$

The sensitivity of the global stiffness matrix to the height variables $\partial \mathbf{K}/\partial \bar{h}_{ij}$ is obtained by assembling the sensitivity of the element matrix $\partial \mathbf{k}(\rho_k)/\partial \bar{h}_{ij}$. By using the chain rule, the term $\partial \mathbf{k}(\rho_k)/\partial \bar{h}_{ij}$ can be calculated as

$$\frac{\partial \mathbf{k}(\rho_k)}{\partial \bar{h}_{ij}} = \frac{\partial \rho_k}{\partial h_c}\frac{\partial h_c}{\partial \bar{h}_{ij}}\mathbf{k}_k = \frac{\beta e^{\beta(h_c-(H_k+H_{k-1})/2)}}{(1+e^{\beta(h_c-(H_k+H_{k-1})/2)})^2} \cdot N_{i,p}(\xi) \cdot N_{j,q}(\eta) \cdot \mathbf{k}_k \tag{24}$$

where, $\mathbf{k}_k$ is the stiffness matrix of the $k$-th element. Similarly, the sensitivity of volume constraint can be expressed as

$$\frac{\partial g(\bar{h})}{\partial \bar{h}_{ij}} = \frac{1}{V_0}\sum_{k=1}^{n_e}\frac{\partial \rho_k}{\partial \bar{h}_{ij}}V_k = \frac{1}{V_0}\sum_{k=1}^{n_e}\frac{\partial \rho_k}{\partial h_c}\frac{\partial h_c}{\partial \bar{h}_{ij}}V_k \tag{25}$$

where $V_k$ and $V_0$ denote the element volume and the total volume of design domain, respectively.

### 5.3 Optimization procedure

The procedure is shown in Fig. 10. First, 3D triangular meshing is fulfilled for the surface model. Mesh parameterization is further carried out by means of the energy minimization approach and/or the seam cutting approach depending upon the surface closeness. The mesh distortion is then evaluated to check the quality of mesh parameterization for the satisfaction of the allowable range. If the mesh distortion is beyond the allowable range, the seam cutting approach is applied. If the mesh distortion is still beyond the allowable range, the multi-chart segmentations approach is needed until the satisfaction of the allowable range. Thereafter, triangular prism finite elements are generated based on the 3D meshed surface and controlled by the maximum stiffener height. A B-spline height field is then defined over the parametric domain. Finite element analysis and sensitivity analysis are carried out and height design variables are finally optimized. It is important to emphasize that in practice, thin-walled structures usually need to ensure a smooth outer surface. Stiffeners are usually placed on the other side of the skin. However, this placement is not convenient for observing the stiffener layout. Hence, we plot the stiffeners on the outer surface.

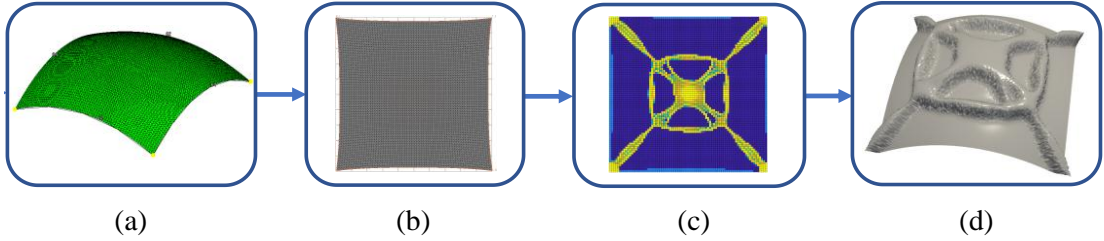

(a)          (b)          (c)          (d)

Fig. 10 Main procedure of stiffener layout optimization. (a) Meshed surface. (b) Mesh parameterization. (c) Control points distribution. (d) Stiffener layout.

## 6 Numerical examples

### 6.1 Hollow cylinder

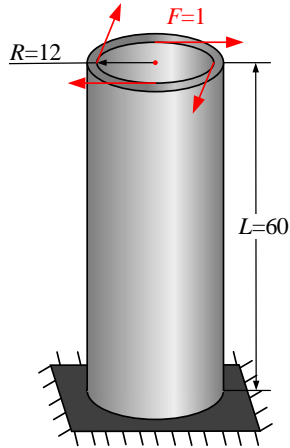

Fig. 11 Geometric model of a hollow cylinder.

The first example concerns a hollow cylinder loaded in torsion by four concentrated forces, as depicted in Fig. 11. It was studied in our previous work (Feng et al. 2021). As this problem has a regular surface, its parametric equation is analytically available. Hence, the analytical parametric mapping can be obtained as the reference result to verify the effectiveness of the mesh parameterization. Here, the inner radius and length of the hollow cylinder are 12 mm and 60 mm, respectively. Material properties are: Young's modulus E = 2.1 $\times 10^5$ Pa and Poisson's ratio $v$ = 0.3. The allowable maximum stiffener height is $H_{max}$ = 0.5 mm. The degrees of B-spline height field $h$ in $\xi$-direction and $\eta$-direction are set to be $p = q = 3$. The volume fraction is constrained below 10% for the design domain. The cylindrical surface is represented by 8484 triangles. 3-layer meshes are used along the stiffener height direction with a total number of 25452 triangular prism elements and 16968 nodes for finite element analysis.

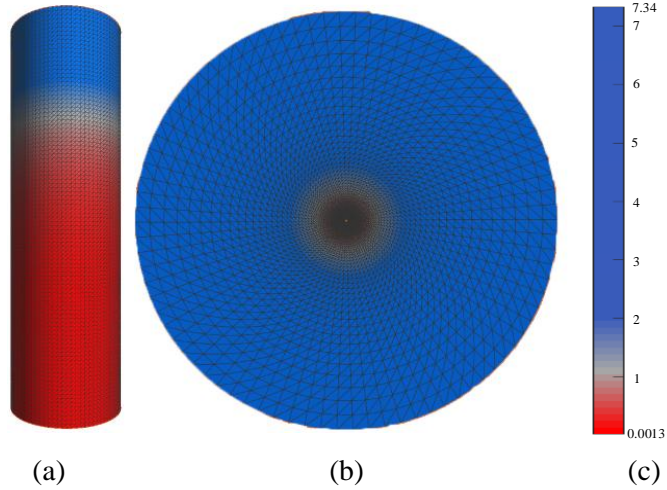

(a) (b) (c)

Fig. 12 Mesh parameterization with severely compressed distortion by the ARAP.

Geometrically, this is a multi-boundary problem, the mesh on the cylindrical surface is flattened onto a 2D circular ring with the help of energy minimization approach. In Fig. 12(b), red and blue colors indicate that the parameterized mesh is compressed and stretched after being flattened. The distortion range is [0.0013, 7.34] that severely exceeds the range defined in Section 4. In detail, only less than one percent of the B-spline control points are distributed in the small region but affect nearly two-thirds of the cylinder. As a result, only one small portion of the cylinder shown in Fig. 13(a) is optimized for stiffener layout. The colors in Fig. 13(b) indicate the values related to the control points, yellow indicates the maximum value $H_{max}$, blue indicates the minimum value, and the other colors indicate the transition values. The non-optimized part is a large portion in the physical domain but is compressed into a very small region in the parametric domain with strong mesh distortion. This mismatch between the uniform B-spline parameterization and the non-uniform mesh parameterization reveals why the stiffener layout cannot be fully optimized over the meshed surface.

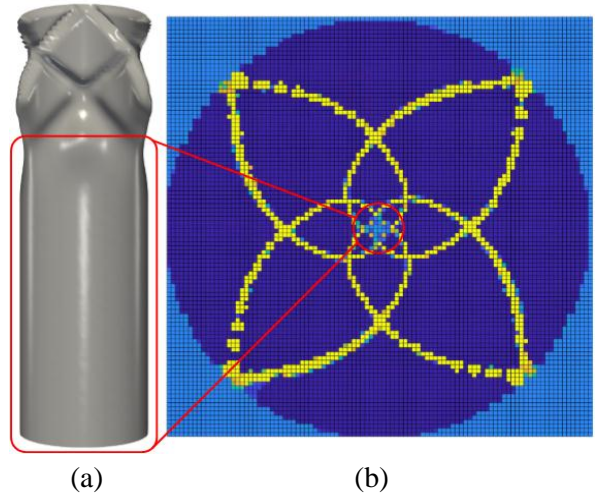

(a) (b)

Fig. 13 Locally optimized stiffener layout due to compressed mesh distortion.

To overcome this difficulty, the seam cutting approach is adopted for mesh parameterization. Fig. 14(b) shows the flattened mesh in a $1 \times 1$ square that perfectly fits the B-spline parametric domain $(\xi,\eta) \in [0, 1] \times [0, 1]$. Here, $100 \times 100$ B-spline control points are defined as design variables in the B-spline parametric domain. Fig. 15 plots the optimization results of stiffener layout obtained by analytical parameterization and the seam cutting processing for the energy minimization approach, respectively. Results are almost the same with crossing stiffeners in an inclined way to resist the torsion load as much as possible. Fig. 16 depicts the convergence curves. Both the objective function and volume fraction gradually decrease and are then stabilized along with the iteration.

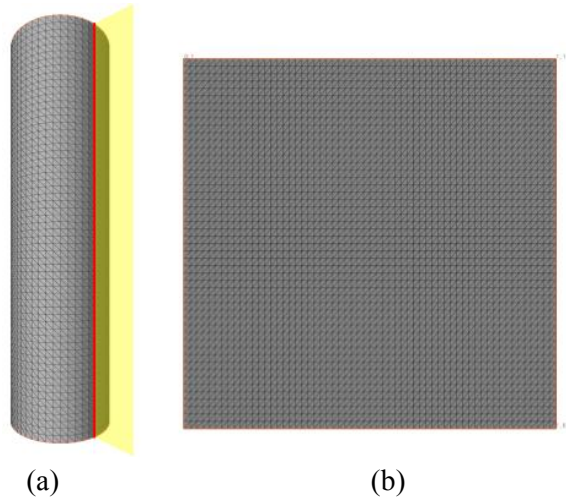

(a) (b)

Fig. 14 (a) 3D meshed surface of the hollow cylinder using the seam cutting approach. (b) Mesh parameterization results.

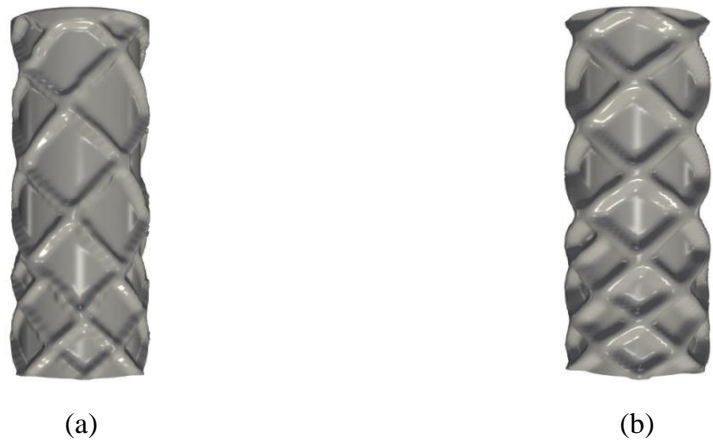

(a) (b)

Fig. 15 Comparison of stiffener layout. (a) Analytical parameterization. (b) Seam cutting approach combined with the ARAP.

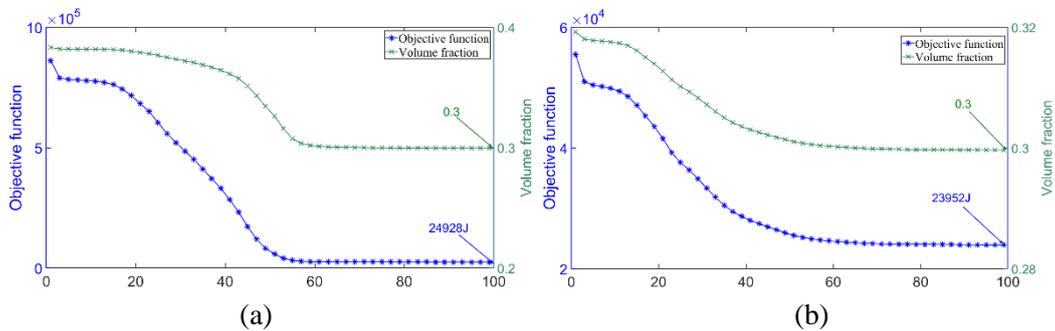

(a) (b)

Fig. 16 Convergence curves. (a) Analytical parameterization. (b) Mesh parameterization.

## 6.2 Aircraft front fuselage model

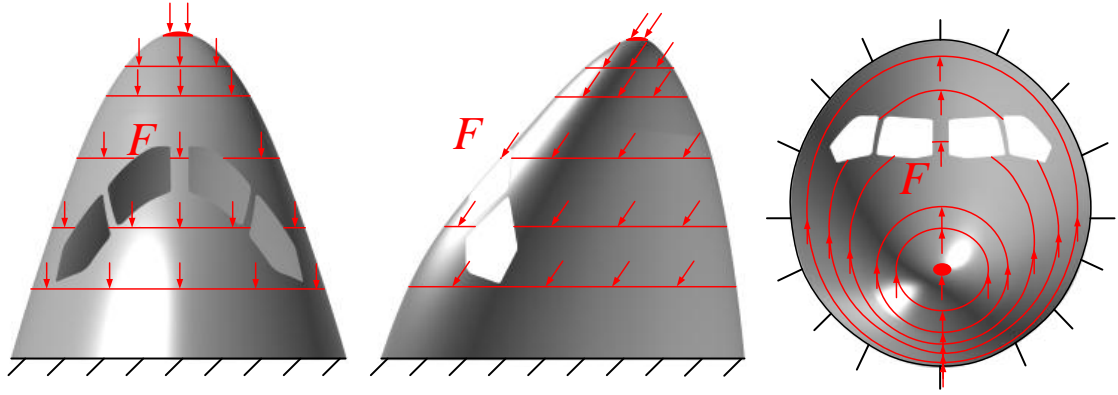

Fig. 17 Geometric model of an aircraft front fuselage model.

Fig. 17 shows the model of an aircraft front fuselage structure. It is fixed at the rear side and the aerodynamic loads are simplified and modeled as section loads according to the work of Zhu et al. (2016). Young's modulus and Poisson's ratio are $E = 9 \times 10^{10}$ Pa and $v = 0.33$, respectively. Here, suppose the B-spline degrees of height field $h$ are set to be $p = q = 5$. A number of $200 \times 200$ control points are employed for optimization. A volume fraction of 10% is imposed to constrain the design. The allowable maximum stiffener height is $H_{max} = 10$ mm.

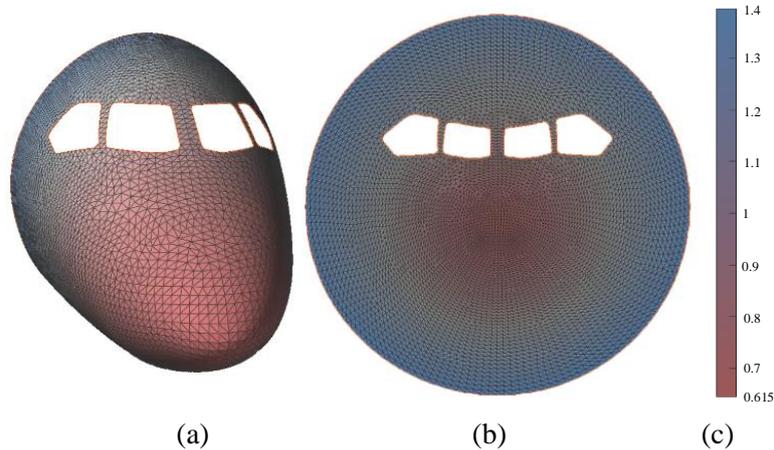

(a)        (b)        (c)

Fig. 18 Mesh parameterization and mesh distortion of the aircraft front fuselage model. (a) Aircraft front fuselage model. (b) Mesh parameterization result. (c) Mesh distortion scale.

Here, the aircraft front fuselage is geometrically represented by 12084 triangular meshes, as shown in Fig. 18(a) and the FE model consists of 24168 triangular prism elements. The energy minimization approach (ARAP) is first adopted to carry out mesh parameterization. Fig. 18(b) shows the mesh parameterization of the aircraft front fuselage surface. The mesh distortion represented in color on the 3D mesh and flattened mesh satisfies the allowable range. Fig. 19(a) plots the distribution of control points on the parametric domain. Fig. 19(b) and (c) depict the optimized stiffener layout in different views. The structure is reinforced around sections where loads are applied. The main force transmission stiffeners are outlined to arrive at the fixed end.

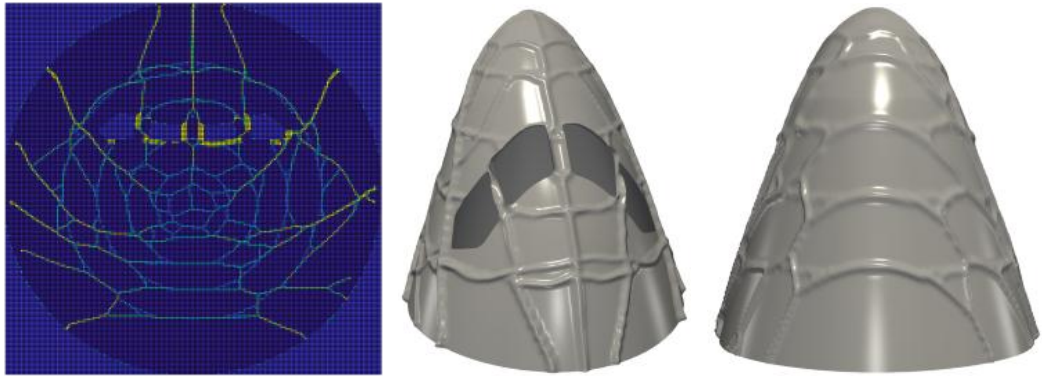

(a) (b) (c)

Fig. 19 Optimized stiffener layout and control point distribution.

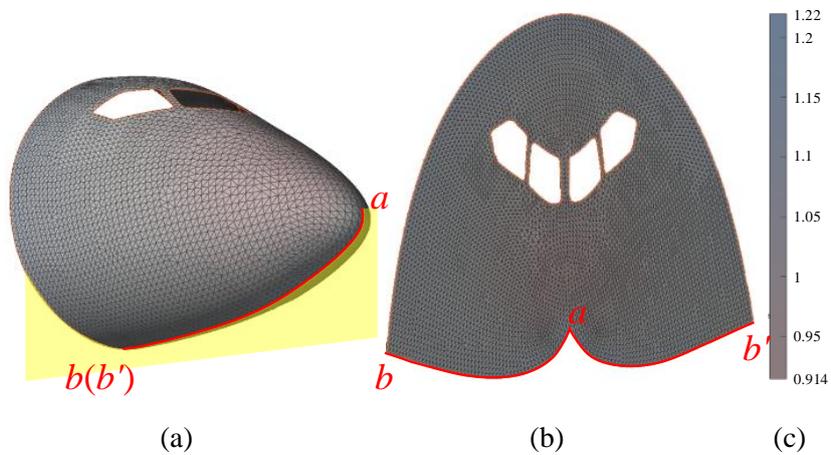

(a) (b) (c)

Fig. 20 Mesh parameterization by the seam cutting approach. (a) Aircraft front fuselage model with the seam cutting. (b) Mesh parameterization result. (c) Mesh distortion scale.

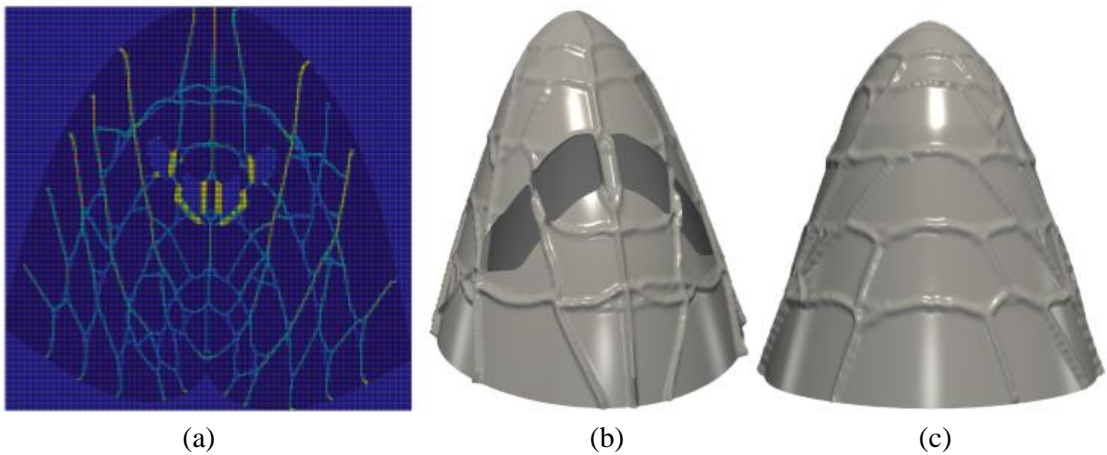

(a) (b) (c)

Fig. 21 Optimized stiffener layouts and control point distribution by using the cutting approach.

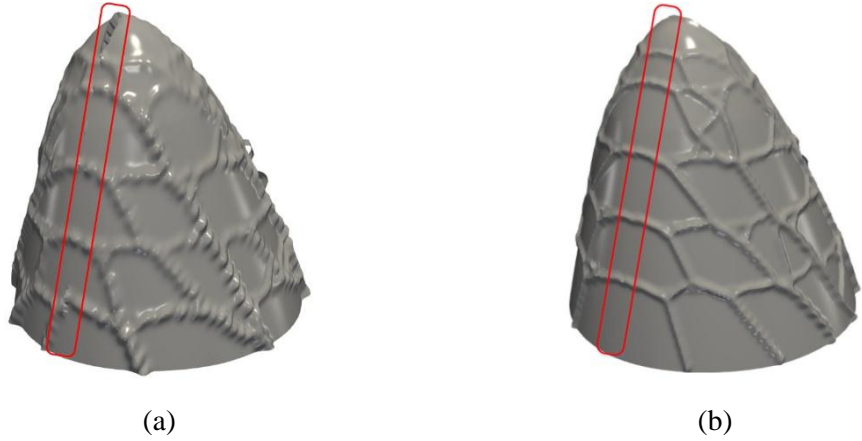

(a)            (b)

Fig. 22 Reduction of the effect of defects at the cut. (a) Number of elements is 8372 and the B-spline degrees are $p = q = 3$. (b) Number of elements is 24168 and the B-spline degrees are $p = q = 5$.

The seam cutting approach is further tested. Fig. 20(a) shows the 3D meshed surface with the seam cutting. Fig. 20(b) indicates that the maximum mesh distortion is reduced to 1.22 and perfectly satisfies the allowable range. Fig. 21 depicts the optimized result in different views. Clearly, with the preprocessing of the seam cutting approach, the optimized result is almost the same as the direct use of the energy minimization approach. The introduction of seams inevitably leads to the presence of discontinuities in the vicinity of the seams, but this defect can be compensated by adjusting the B-spline degree, the number of elements, the number of control points. As shown in Fig. 22(a), when the number of elements and the B-spline degree are small, there is an obvious discontinuity of stiffeners at the seams, while when the number of elements and the B-spline degree are increased, the area affected by the seams decreases and the influence of the higher order continuity of the B-spline increases, and in Fig. 22(b), the discontinuity is barely visible at the seams. The degree of the B-spline, the number of control points and the number of elements can affect the filtering radius for the avoidance of jagging boundaries and checkerboard problems, and can also influence the transition interval of the stiffened structure. More details can refer to the works of Feng et al. (2021) and Qian (2013). Fig. 23 shows convergence curves with and without cutting. Finally, the structural compliances are 100.3 J and 98.1 J, respectively.

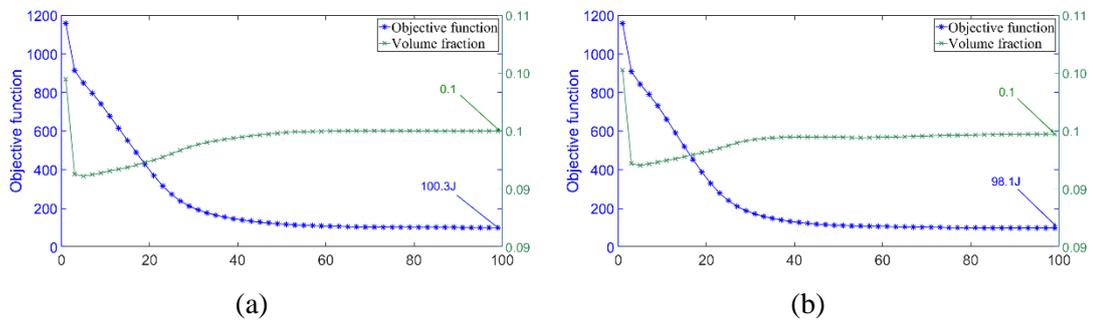

(a)            (b)

Fig. 23 Convergence curves. (a) The cutting approach. (b) The energy minimization approach.

## 6.3 Space capsule

This is a structure with a closed surface and also a non-developable surface. Fig. 24 shows the simplified model. Suppose the structure is fixed at location $P$. A distributed load is applied vertically on the structural cross-section. Young's modulus and Poisson's ratio are $E = 9 \times 10^{10}$ Pa and $v = 0.3$, respectively. Likewise, the B-spline degrees are set as $p = q = 5$ and the number of control points is $200 \times 200$. The space capsule is meshed by 22286 triangles and the corresponding FE model contains 44572 triangular prism elements.

The seam cutting approach is used and four different cutting cases are studied in Fig. 25. It can be seen that the maximum mesh distortions are almost the same and the optimized results have slight differences. The structural compliance is 3485.2 J, 3416.9 J, 3276.1 J from (a) to (c) in order. From the magnitude of structural compliance, it can be seen that the optimization results are best when the mesh distortion and the seam are small. The comparison of cases (a-c) indicates that under the same mesh quality control, the shorter the seam, the better the optimized result.

Furthermore, the multi-chart segmentations approach is also applied to split the surface into two parts. The unfolding results of the multi-chart segmentations approach and the corresponding optimized stiffener layout are given in Fig. 25(d). Compared to the seam cutting approach, the distortion obtained with the multi-chart segmentations approach is significantly small with the value being [0.632, 1.17]. As shown in Fig. 26, the optimized result holds the same stiffener layout with the structural compliance being 3368.3 J.

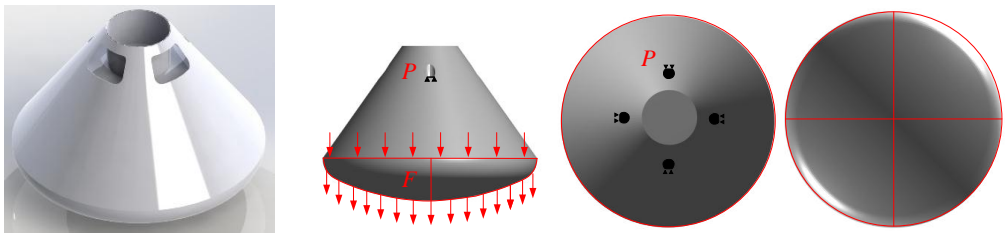

Fig. 24 Space capsule model and load conditions.

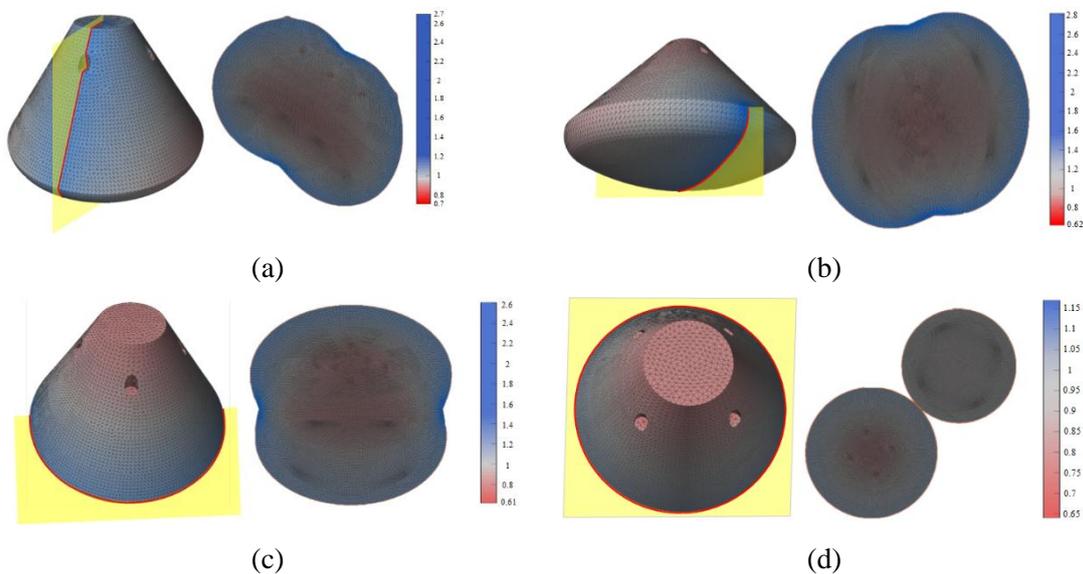

(a)  (b)

(c)  (d)

Fig. 25 Mesh parameterization of the model with different cutting approaches.

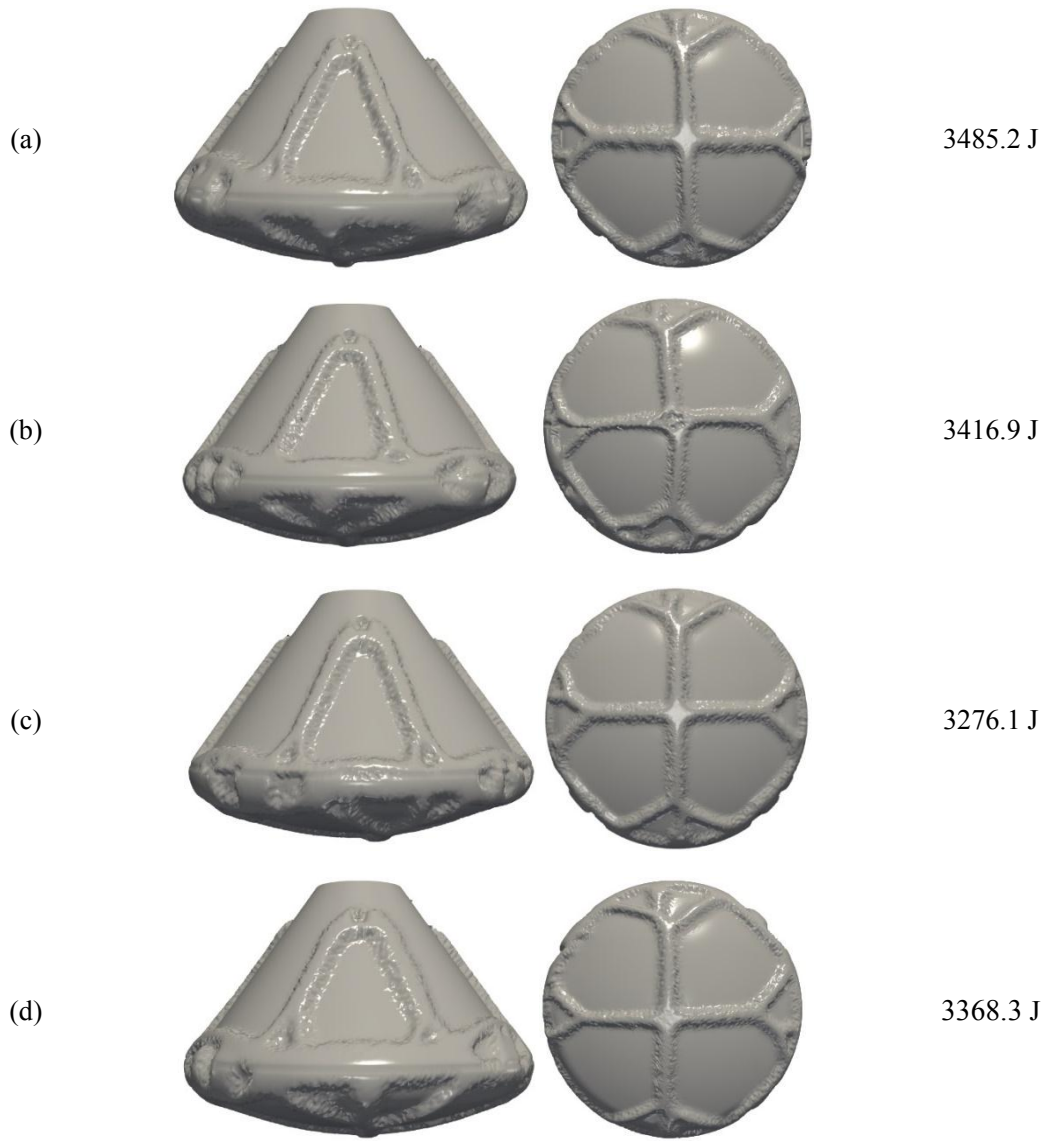

(a) 3485.2 J

(b) 3416.9 J

(c) 3276.1 J

(d) 3368.3 J

Fig. 26 Optimized stiffener layouts and structural compliance with different cutting approaches.

## 7 Conclusions

In this paper, the mesh parameterization is combined with the B-spline parameterization to deal with stiffener layout optimization of thin-walled structures with complex surfaces. Different mesh parameterization approaches are presented and the matching issue of the combined parameterization is studied. The mesh distortion is evaluated to check the quality of mesh parameterization from the physical domain to the parametric domain and an allowable range is proposed.

Based on numerical examples, the energy minimization approach and the cutting processing of complex surfaces are investigated to show their effects upon optimized results. The mismatch between the mesh parameterization and B-spline parameterization is highlighted with the help of the first numerical example. The last two examples show that the combined parameterization method is also effective for both closed surfaces and complex surfaces with cutouts. Although the current work is focused on compliance minimization problem, it can naturally be extended to other kinds of design problems such as buckling, vibration and thermo-mechanical designs for stiffener layout optimization.

# 8 Replication of results

All the results and datasets in this paper are generated using our in-house MATLAB codes. The source codes can be available only for academic use from the corresponding author with a reasonable request.

**Acknowledgments** This work is supported by National Natural Science Foundation of China (12032018).

**Compliance with ethical standards**

**Conflict of interest** The authors declare that they have no conflict of interest.

**Reference**

Afonso S, Sienz J, Belblidia F (2005). Structural optimization strategies for simple and integrally stiffened plates and shells. Engineering Computations, 22(4), 429-452.

Bendsøe M (1989). Optimal shape design as a material distribution problem. Structural and Multidisciplinary Optimization, 1(4):193-202.

Bennis C, Vézien J, Iglésias G (1991). Piecewise surface flattening for non-distorted texture mapping. ACM SIGGRAPH Computer Graphics, 25(4), 237-246.

Cheng K, Olhoff N (1981). An investigation concerning optimal design of solid elastic plates. International Journal of Solids and Structures, 17(3):305–323.

Cheng K, Olhoff N (1982). Regularized formulation for optimal design of axisymmetric plates. International Journal of Solids and Structures, 18(2):153–169.

Chung J, Lee K (1997). Optimal design of rib structures using the topology optimization technique. In Proceedings of the Institution of Mechanical Engineers, Part C: Journal of Mechanical Engineering Science, 211(6):425–437.

Ding X, Yamazaki K (2004). Stiffener layout design for plate structures by growing and branching tree model (application to vibration-proof design). Structural and Multidisciplinary Optimization, 26(1-2):99–110.

Eck M, Rose T, Duchamp T, Hoppe H, Lounsbery M, Stuetzle W (1995). Multiresolution analysis of arbitrary meshes. Springer Berlin Heidelberg, 173-182.

Feng S, Zhang W, Meng L, Xu Z, Chen L (2021). Stiffener layout optimization of shell structures with B-spline parameterization method. Structural and Multidisciplinary Optimization, 63:2637-2651.

Floater M (1997). Parametrization and smooth approximation of surface triangulations. Computer Aided Geometric Design, 14(3), 231-250.

Hao P, Wang B, Tian K, Li G, Du K, Niu F (2016). Efficient optimization of cylindrical stiffened shells with reinforced cutouts by curvilinear stiffeners. AIAA Journal, 54(4):1350–1363.

Hormann K, Lévy B, Sheffer A (2007). Mesh parameterization: Theory and practice. In: Proc. of the SIGGRAPH Asia ACM SIGGRAPH ASIA Courses. New York: ACM Press, 12:1−12:87.

Hou J, Zhu J, He F, Zhang W, Guo W (2017). Stiffeners layout design of thin-walled structures with constraints on multi-fastener joint loads. Chinese Journal of Aeronautics, 30(4):1441–1450.

Ji J, Ding X, Xiong M (2014). Optimal stiffener layout of plate/shell structures by bionic growth method. Computers & Structures, 135:88–99.

Kiendl J, Schmidt R, Wüchner R, Bletzinger K (2014). Isogeometric shape optimization of shells using semi-analytical sensitivity analysis and sensitivity weighting. Computer Methods in Applied Mechanics and Engineering, 274, 148-167.

Kang P, Youn S (2016). Isogeometric shape optimization of trimmed shell structures. Structural and Multidisciplinary Optimization, 53(4), 825-845.

Lévy B, Petitjean S, Ray N, Maillot J (2002). Least squares conformal maps for automatic texture atlas generation. ACM Transactions On Graphics (TOG), 21(3), 362-371.

Liang K, Yang C, Sun Q (2020). A smeared stiffener based reduced-order modelling method for buckling analysis of isogrid-stiffened cylinder. Applied Mathematical Modelling, 77, 756-772.

Liu L, Zhang L, Xu Y, Gotsman C, Gortler S (2008). A local/global approach to mesh parameterization, Computer Graphics Forum, 27(5):1495-1504.

Liu S, Li Q, Chen W, Hu R, Tong L (2015). H-dgtpa heaviside-function based directional growth topology parameterization for design optimization of stiffener layout and height of thin-walled structures. Structural and Multidisciplinary Optimization, 52(5):903–913.

Maillot J, Yahia H, Verroust A (1993). Interactive texture mapping. In Proceedings of the 20th annual conference on Computer graphics and interactive techniques, pp. 27-34.

Pinkall U, Polthier K (1993). Computing discrete minimal surfaces and their conjugates. Experimental Mathematics, 2(1), 15-36.

Qian X (2013) Topology optimization in B-spline space. Computer Methods in Applied Mechanics and Engineering, 265:15–35.

Seo Y, Kim H, Youn S (2010). Isogeometric topology optimization using trimmed spline surfaces. Computer Methods in Applied Mechanics and Engineering, 199(49-52), 3270-3296.

Sander P, Snyder J, Gortler S, Hoppe H (2001). Texture mapping progressive meshes. In Proceedings of the 28th annual conference on Computer graphics and interactive techniques, pp. 409-416.

Sander P, Justine W, Gortler S, Snyder J, Hoppe H (2003). Multi-chart geometry images. In Proceedings of the 1st Symposium on Geometry Processing, pp. 138–145.

Sethian J, Wiegmann A (2000). Structural boundary design via level set and immersed interface methods. Journal of Computational Physics, 163(2):489–528.

Sheffer A, Sturler E (2000). Surface parameterization for meshing by triangulation flattening. In Proceedings of the 9th International Meshing Roundtable, Sandia National Laboratories, pp. 161–172.

Sheffer A, Praun E, Rose K. (2006). Mesh parameterization methods and their applications. Foundations and Trends in Computer Graphics and Vision, 2(2), 105-171.

Sorkine O, Cohen-Or D, Goldenthal R, Lischinski D (2002). Bounded-distortion piecewise

mesh parameterization. In Proceedings of IEEE Visualization 2002, pp. 355–362.

Tian K, Li H, Huang L, Huang H, Zhao H, Wang B (2020). Data-driven modelling and optimization of stiffeners on undevelopable curved surfaces. Structural and Multidisciplinary Optimization, 62(6), 3249-3269.

Tutte W (1960). Convex representations of graphs. In Proceedings of the London Mathematical Society, 3(1), 304-320.

Wang D, Zhang W. (2012). A bispace parameterization method for shape optimization of thin-walled curved shell structures with openings. International Journal for Numerical Methods in Engineering, 90(13), 1598-1617.

Wang D, Abdalla M, Zhang W (2017). Buckling optimization design of curved stiffeners for grid-stiffened composite structures. Composite Structures, 159, 656-666.

Wang D, Abdalla M, Wang Z, Su Z (2019). Streamline stiffener path optimization (sspo) for embedded stiffener layout design of non-uniform curved grid-stiffened composite (ncgc) structures. Computer Methods in Applied Mechanics and Engineering, 344:1021–1050.

Wang D, Yeo S, Su Z, Wang Z, Abdalla M (2020). Data-driven streamline stiffener path optimization (SSPO) for sparse stiffener layout design of non-uniform curved grid-stiffened composite (NCGC) structures. Computer Methods in Applied Mechanics and Engineering, 365, 113001.

Zhu J, Li Y, Zhang W, Hou J (2016). Shape preserving design with structural topology optimization. Structural & Multidisciplinary Optimization, 53(4), 893-906.